\documentclass[preprint]{aastex}
\usepackage{mkfig}

\begin{document}

\title{Internal Structure of the 2019 April 2 CME}

\author{Brian E. Wood\altaffilmark{1},
  Carlos R. Braga\altaffilmark{2}, Angelos Vourlidas\altaffilmark{3}}
\altaffiltext{1}{Naval Research Laboratory, Space Science Division,
  Washington, DC 20375, USA; brian.wood@nrl.navy.mil}
\altaffiltext{2}{George Mason University, 4400 University Drive, Fairfax, VA
  22030, USA}
\altaffiltext{3}{The Johns Hopkins University Applied Physics Laboratory,
  Laurel, MD 20723, USA}

\begin{abstract}

     We present the first analysis of internal coronal mass ejection (CME)
structure observed very close to the Sun by the Wide-field Imager for Solar
PRobe (WISPR) instrument on board Parker Solar Probe (PSP).  The transient
studied here is a CME observed during PSP's second perihelion passage on
2019 April 2, when PSP was only 40~R$_{\odot}$ from the Sun.  The CME was
also well observed from 1~au by the STEREO-A spacecraft, which tracks
the event all the way from the Sun to 1 au.  However, PSP/WISPR observes internal
structure not apparent in the images from 1~au.  In particular, two linear
features are observed, one bright and one dark.  We model these features as
two loops within the CME flux rope channel.  The loops
can be interpreted as bundles of field lines, with the brightness of
the bright loop indicative of lots of mass being loaded into those field
lines, and with the dark loop being devoid of such mass loading.  It is
possible that these loops are actually representative of two independent
flux rope structures within the overall CME outline.

\end{abstract}


\section{Introduction}

     Launched in 2018, Parker Solar Probe (PSP) is currently exploring the
origins of the solar wind, venturing closer to the Sun than any previous
spacecraft.  This exploration uses primarily a suite of in situ particle and field
instruments, designed to directly measure plasma and field properties.
However, PSP also has an imaging instrument on board, the Wide-field Imager
for Solar PRobe (WISPR) \citep{av16}.  The two heliospheric
imagers that consistitute PSP/WISPR observe white light scattered from the
solar wind in the ram direction of PSP's orbit.

     In addition to exploring the quiescent solar wind, the PSP instruments have
also observed a number of transients that have happened to occur near PSP's
perihelion passages.  In PSP's first perihelion passage, WISPR observed two
noteworthy coronal mass ejections (CMEs).  The first was on 2018 November 1,
characterized by a small circular cavity indicative of a magnetic flux rope (FR)
viewed edge-on \citep{ph20,apr20}.  The second was
a jet-like streamer blob event, which was observed even closer to perihelion
on 2018 November 5.  This event was not only observed by WISPR but also by
coronagraphs on two spacecraft operating at 1 au:  the C2 and C3 constituents
of the Large Angle and Spectrometric COronagraph (LASCO) instrument on board the
SOlar and Heliospheric Observatory (SOHO), and the COR2-A coronagraph on the
Solar TErrestrial RElations Observatory (STEREO).  This provided the first
opportunity for a full morphological reconstruction of a WISPR-observed CME
utilizing three distinct vantage points \citep{bew20}.

     On 2019 April 1-2, during PSP's second perihelion passage, WISPR observed
two larger CMEs than the ones seen the previous orbit \citep{crb21}.  The
events were also observed by STEREO-A and SOHO/LASCO.  The first analyses of
these CMEs focused on measuring their trajectory directions and kinematic
properties, in order to demonstrate consistency between measurements made from
WISPR images and those from 1~au \citep{pcl20,crb21}.
However, the April 2 CME is also noteworthy for having an unusual
appearance in the WISPR images, which \citet{pcl20}
referred to as ``skull-like.''  This curious visage is not apparent in the
images from the 1~au spacecraft, meaning that this is an opportunity to
study CME internal structure with PSP/WISPR that is only visible thanks
to PSP's close proximity to the CME near the Sun.  Although there have already
been morphological reconstructions of overall CME shapes using PSP/WISPR
\citep{apr20,bew20}, this is the first attempt
to reconstruct internal CME structure from WISPR images.

\section{Observations}

\begin{figure}[t]
\plotfiddle{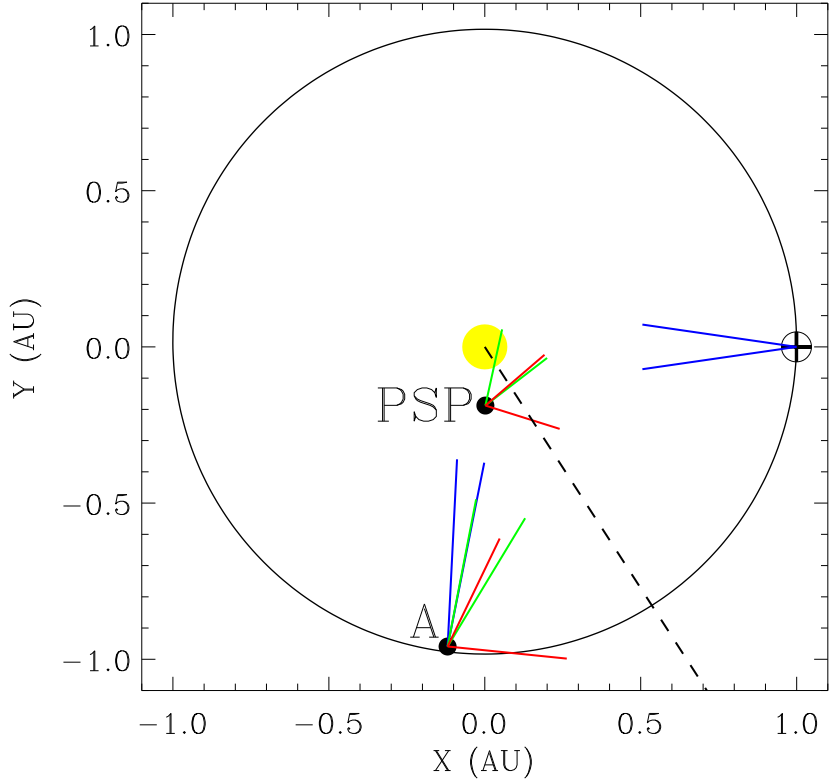}{3.6in}{0}{90}{90}{-280}{-350}
\caption{The positions of Earth, PSP, and STEREO-A in the ecliptic
  plane on 2019~April~2 (in HEE coordinates).  At PSP's position, the green and
  red lines indicate the fields of view of the WISPR-I and WISPR-O detectors.
  At Earth's position, the blue lines indicate the field of view of the LASCO/C3
  coronagraph on SOHO.  At STEREO-A's position, the blue, green, and red lines
  indicate the fields of view of COR2-A, HI1-A, and HI2-A.  The dashed line indicates
  the central trajectory of the 2019 April 2 CME observed by all these imagers.}
\end{figure}
     The second perihelion passage of PSP brought the spacecraft within
35.4~R$_{\odot}$ of the Sun at UT 22:40 on 2019 April 4.  The CME that is
the focus of our study was first seen by the WISPR instrument two days
earlier at about UT 11:00 on 2019 April 2, with PSP 40.0~R$_{\odot}$ from
the Sun.  The viewing geometry is illustrated in Figure~1, which shows the
positions of PSP, STEREO-A, and Earth relative to the CME trajectory, in
heliocentric Earth ecliptic (HEE) coordinates.  The fields of view of six
white light telescopes that image the CME are also shown.  This includes
the C3 coronagraph component of SOHO/LASCO, operating near Earth, which
observes the corona at Sun-center distances in the plane-of-sky 
of $3.7-30$ R$_{\odot}$ \citep{geb95}.  The trajectory of
the 2019 April 2 CME is perfectly placed for tracking
with STEREO-A.  The STEREO-A observations include ones from the COR2-A
coronagraph, observing at angular distances from Sun-center of
$0.7^{\circ}-4.2^{\circ}$ ($2.5-15.6$ R$_{\odot}$), and also the HI1-A and HI2-A
heliospheric imagers, viewing at $3.9^{\circ}-24.1^{\circ}$ and
$19^{\circ}-89^{\circ}$, respectively \citep{rah08,cje09}.
Finally, Figure~1 also depicts the fields of
view of WISPR's two heliospheric imagers:  WISPR-I, imaging at
$13^{\circ}-53^{\circ}$ from the Sun, and WISPR-O, imaging at
$50^{\circ}-108^{\circ}$ \citep{av16}.

     The 2019 April 2 CME probably traces its origins to AR12737, a newly
born active region that emerges at about UT 09:00 on 2019 March 31, as
seen in images from the Atmospheric Imaging Assembly (AIA)
instrument on the Solar Dynamics Observatory (SDO).  This active region
has already received attention for being the likely origin of persistent
Type III radio bursts observed by PSP during its second perihelion passage
\citep{lh21,dhb21}.  Given that the
trajectory inferred for the April 2 CME traces back to near AR12737
\citep{pcl20}, a likely cause of the CME is the destabilization
of the local streamer belt due to the emergence and expansion of AR12737.
However, there is some ambiguity whether it is proper to call AR12737
the CME ``source region,'' as there is little actual surface activity
associated with the eruption.  There is no flare and no filament
eruption, for example.  The only surface activity that might be
directly connected to the CME is a small coronal dimming event apparent
in 171~\AA\ SDO/AIA images, which occurs NW of AR12737 at UT 19:11
on April 1.

\begin{figure}[t]
\plotfiddle{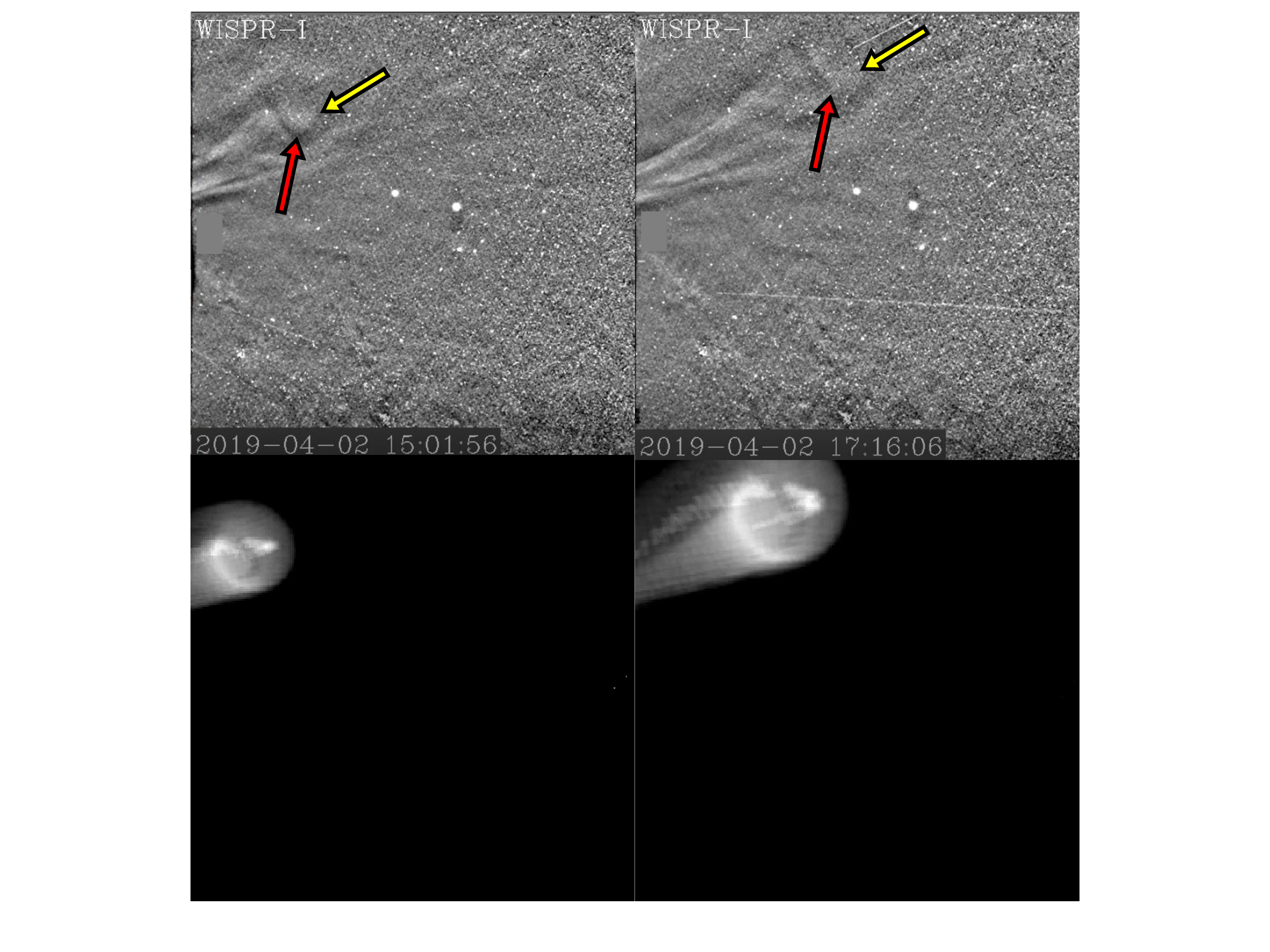}{3.5in}{0}{55}{55}{-195}{-25}
\caption{Two images of the 2019~April~2 CME from the WISPR-I detector on of the
  WISPR instrument on PSP.  Yellow arrows point to the top of a bright internal
  structure that we identify as Loop 1, and red arrows point to a dark internal
  linear structure that we identify as Loop 2 (see Section 4).  Synthetic images
  of the event are shown below the real images, based on the 3-D reconstruction
  described in Section~4.  A movie version of this figure is available online.}
\end{figure}
\begin{figure}[t]
\plotfiddle{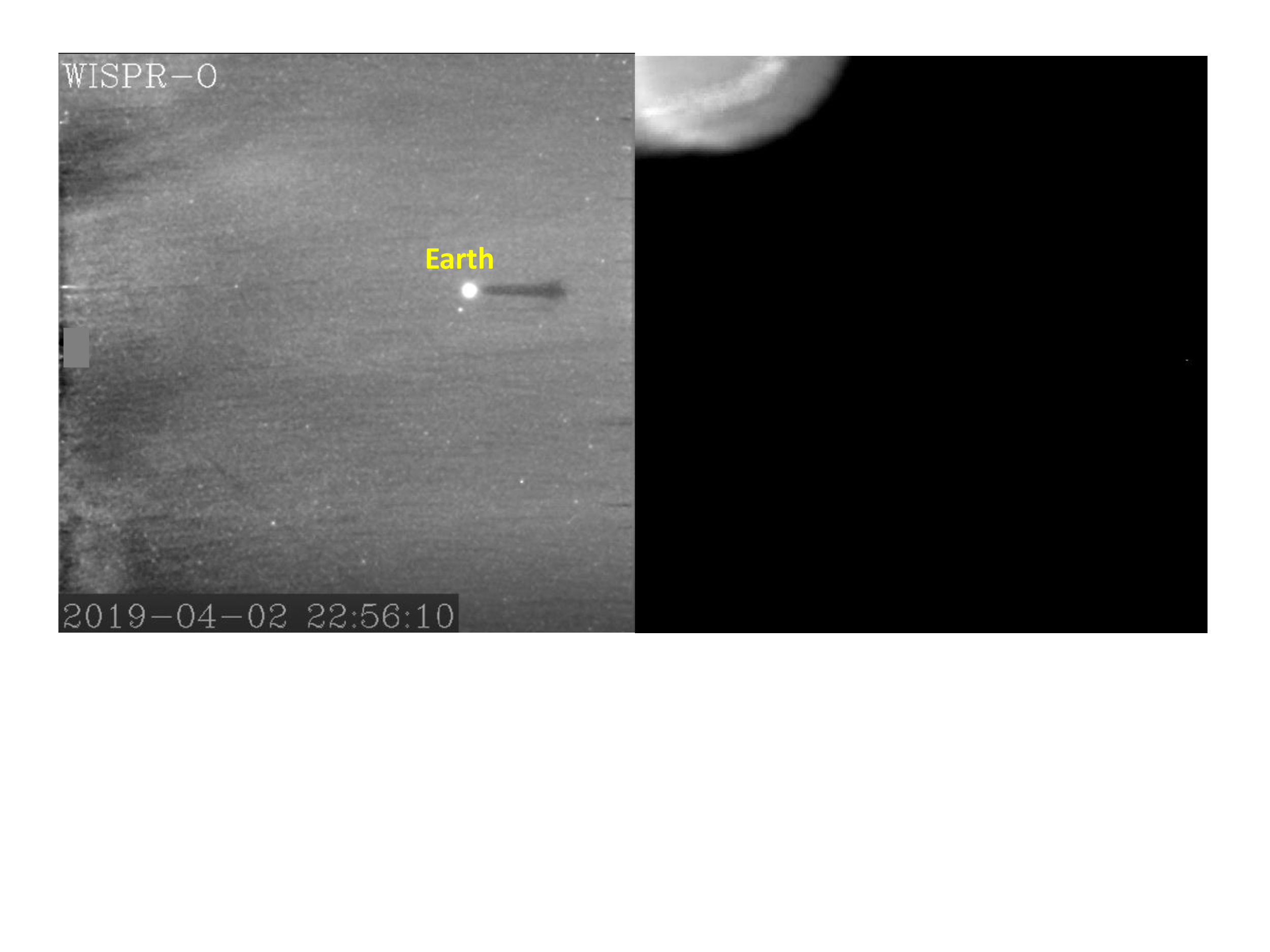}{2.9in}{0}{64}{64}{-230}{-125}
\caption{On the left is a PSP/WISPR-O image of the 2019~April~2 CME, with the
  CME only faintly visible in the upper left corner of the image, and on the
  right is a synthetic image of the event based on the 3-D reconstruction
  described in Section 4.  The CME is far more visible in the movie version
  of this figure, which is available online.}
\end{figure}
\begin{figure}[t]
\plotfiddle{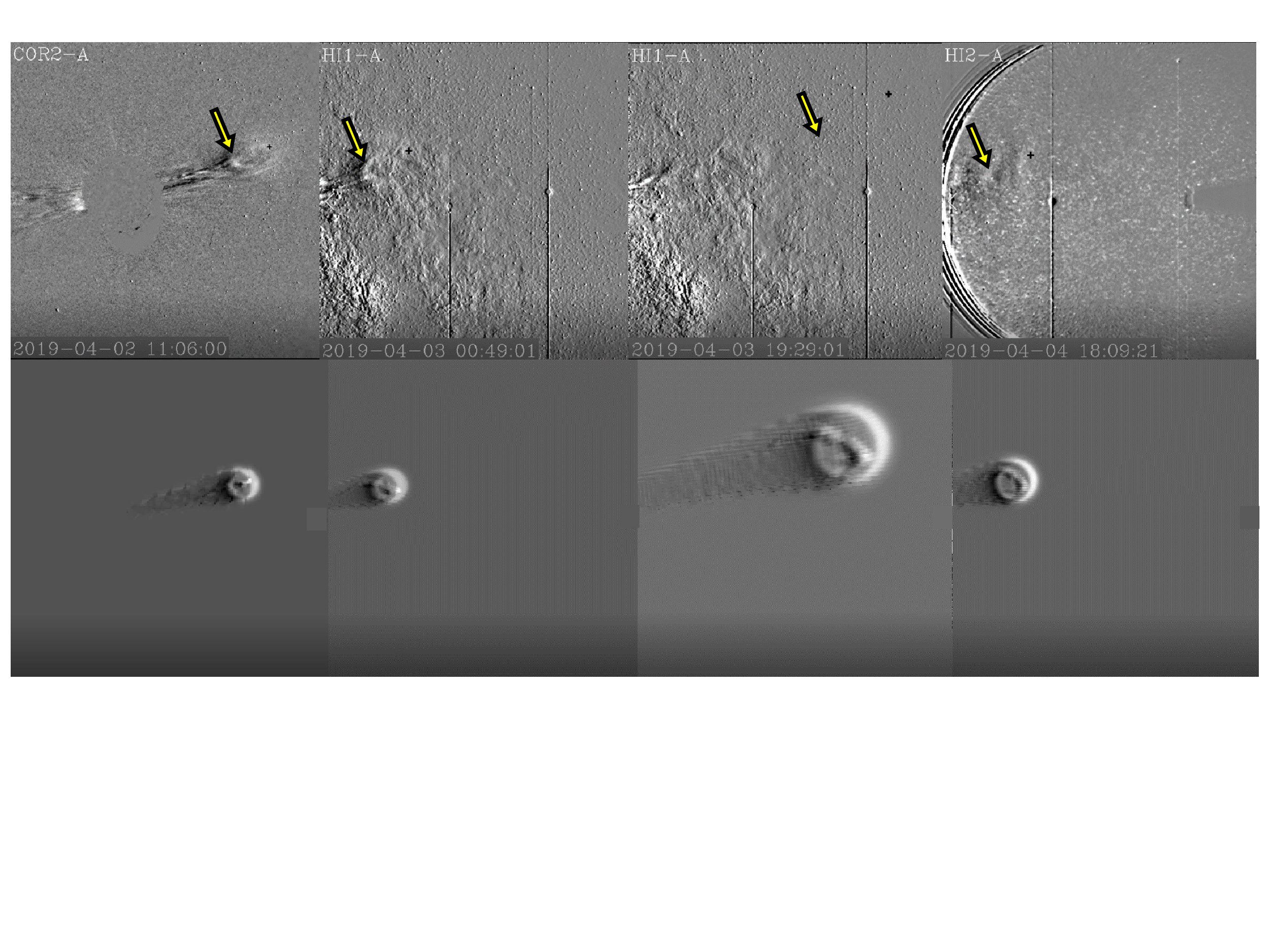}{2.9in}{0}{64}{64}{-230}{-110}
\caption{A sequence of four images of the 2019 April 2 CME from the STEREO-A
  spacecraft, one from COR2-A, two from HI1-A, and one from HI2-A.  A small
  black plus sign marks the position of the CME leading edge as predicted by
  the kinematic model of the CME described in Section 3.  Yellow arrows point
  to a bright pileup region at the back end of the CME flux rope channel caused
  by a fast ambient wind coming from behind.  Synthetic images of the event
  are shown below the real images, based on the 3-D reconstruction
  described in Section~4.}
\end{figure}
\begin{figure}[t]
\plotfiddle{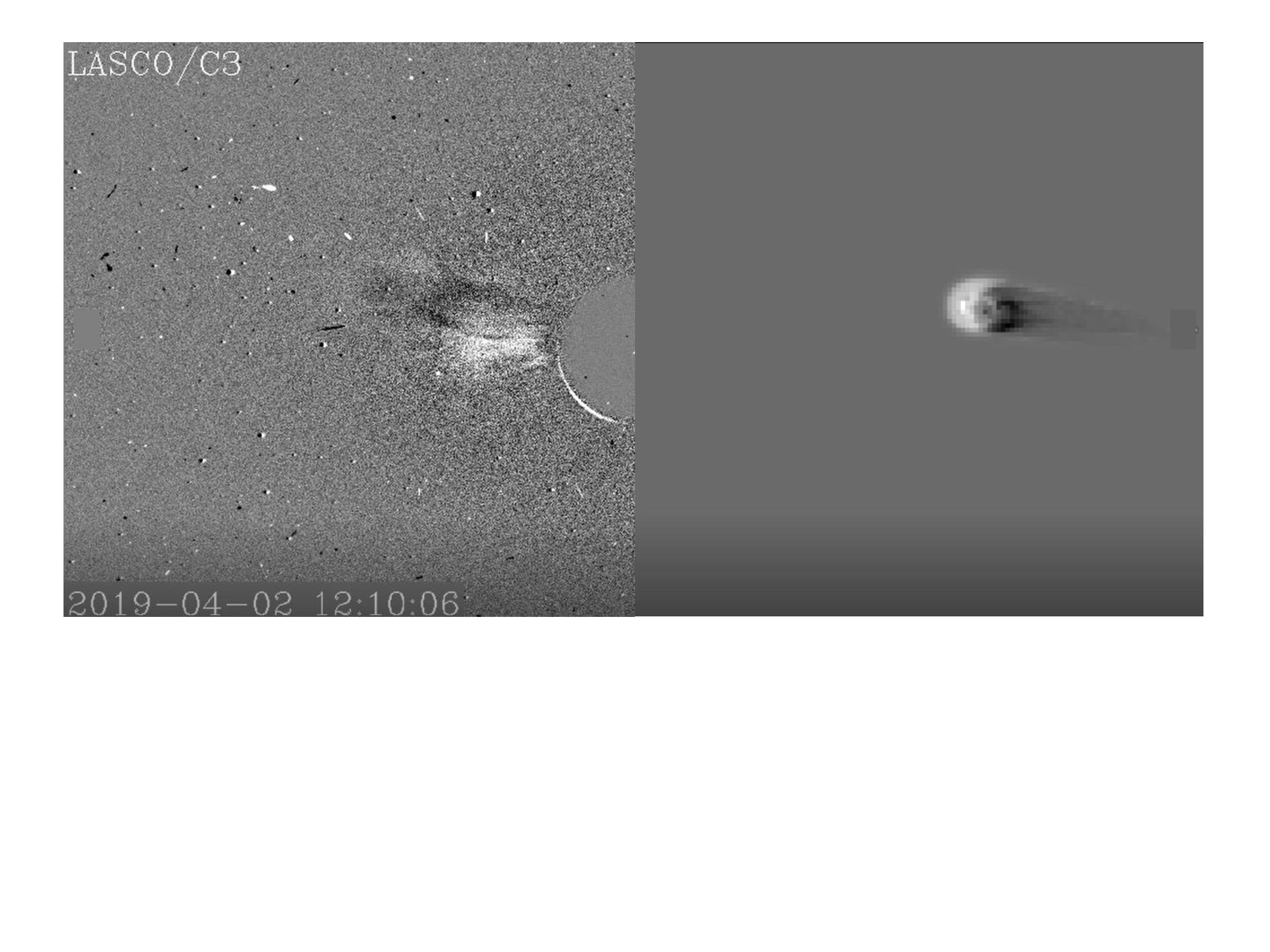}{2.9in}{0}{64}{64}{-230}{-125}
\caption{On the left is a LASCO/C3 image of the 2019~April~2 CME, and on the
  right is a synthetic image of the event based on the 3-D reconstruction
  described in Section 4.  A movie version of this figure is available online.}
\end{figure}
     Figures 2-5 show a selection of images of the CME, from all six of
the aforementioned white light telescopes.  The WISPR images in Figures 2-3
are shown after subtraction of an average image, in order to remove static
structure and focus attention on the transient emission.  The STEREO-A and
LASCO/C3 images in Figures 4-5 are running difference images, with the prior
image subtracted from each image, once again to emphasize transient emission.
It is worth noting that each of Figures 2-5 are accompanied by movie versions
available in the online article, which display the CME far
better and more comprehensively than static images can.

     The details of the CME are best seen in the WISPR-I images in Figure 2.
The outline of the top of the CME is a roughly circular cavity suggestive of
an FR channel observed edge-on.  Inside the cavity is a bright, linear
feature, which stretches back toward the Sun, the top of which is marked by
a yellow arrow in Figure 2.  Even more unusual is a dark, linear feature
superposed onto the bright feature, oriented perpendicular to the CME's
propagation direction, marked by a red arrow in Figure 2.  It is the
combination of these two features that give the CME what \citet{pcl20}
describe as a ``skull-like'' appearance.  Interpreting this internal structure
is the primary goal of our study.

     This internal structure is only apparent in the WISPR-I images.
In the images from 1 au (see Figures 4-5), the CME is simply too faint and
far away to discern this internal structure.  It should be noted that if
we were using those data to study internal structure, we would not be
using the running difference technique as we are in Figures 4-5, as the imprint
of the shadow of the previous image on each image confuses the internal
appearance.  In WISPR-O data, the CME skims
the top of the field of view (see Figure 3), and only the bottom edge of the
CME is really seen.  In the analysis described below, the WISPR-O data are
nevertheless useful for constraining the proximity of that part of the CME to
PSP, as the apparent speed of the CME through the WISPR-O field of view is
a sensitive diagnostic to its proximity, with a closer proximity leading to
a faster apparent propagation.  The WISPR-O data are also notable for having
Earth in the field of view, meaning that this is the first reported instance
of a telescope near the Sun looking back toward Earth and seeing a CME
propagating through the interplanetary space in between, as illustrated
in Figure 1.

\section{Kinematics}

     Our primary goal is to model the internal structure of the 2019 April 2
CME, but there are several intermediate steps on the way to that final goal.
One is to have a full 3-D morphological model of the overall CME outline,
within which the internal structure can be placed.  Another is to have a
full kinematic model for the CME, so that we can properly match the
appearance of the CME in all the telescope fields of view at all times.

     The CME kinematics have already been extensively studied by \citet{pcl20}
and \citet{crb21}.  However, these analyses
focused on the use of this CME to test techniques for measuring CME kinematics
from WISPR images.  We present here a kinematic model of the event
applicable for the full time range of available observations.  It is actually
STEREO-A that is by far the best platform for studying this CME's kinematics,
as STEREO-A tracks the event continuously from near the Sun all the way to
1 au.  Our kinematic analysis therefore focuses on STEREO-A data.

     From COR2-A, HI1-A, and HI2-A images, we measure
the elongation angle, $\epsilon$, of the leading edge of the CME as a function
of time.  Converting $\epsilon$ to actual distances from Sun-center, $r$,
requires assumptions about the shape of the CME front.  Consistent with
past analyses \citep{bew17,bew20}, we use the so-called
``harmonic mean'' approximation from \citet{nl09},
\begin{equation}
r=\frac{2d\sin \epsilon}{1+\sin(\epsilon+\phi)},
\end{equation}
where $d$ is the distance from STEREO-A to the Sun and $\phi$ is
the angle between the CME's central trajectory and the STEREO-A/Sun line.
This essentially approximates the CME as a sphere centered
halfway between the Sun and the CME's leading edge.  Based on the
morphological analysis described in the next section, we infer a trajectory
direction of $\phi=41^{\circ}$ relative to STEREO-A, and the measured
elongation angles then translate to the CME leading edge distances shown
in the top panel of Figure 6.

\begin{figure}[t]
\plotfiddle{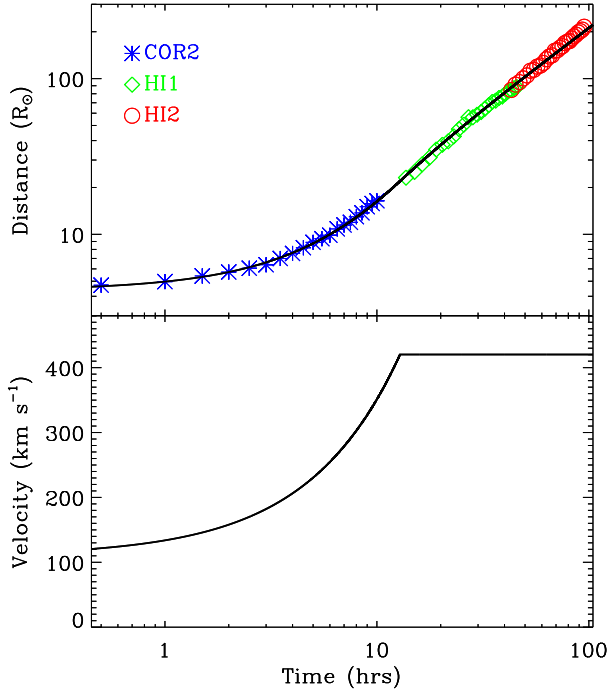}{3.1in}{0}{75}{75}{-220}{-285}
\caption{The top panel shows distance measurements for the leading
  edge of the 2019 April 2 CME as a function of time based on
  images from STEREO-A, specifically COR2-A (blue asterisks), HI1-A
  (green diamonds), and HI2-A (red circles).
  The $t=0$ point on the time axis corresponds to UT 03:06 on
  April 2.  These data points are fitted with a simple kinematic
  model assuming a constant acceleration phase followed by a constant
  velocity phase.  The solid line is the best fit, and the bottom
  panel shows the inferred velocity profile.}
\end{figure}
     In order to infer a velocity profile for the CME, we fit a simple
two-phase kinematic model to the data, assuming a
phase of constant acceleration followed by a phase of constant
velocity.  In the resulting fit in Figure 6, the CME accelerates at a
rate of 6.7 m~s$^{-2}$ for about 13 hours before leveling out at a
final speed of 420 km~s$^{-1}$.  These acceleration and speed measurements
are in excellent agreement with the $6.667\pm 2.714$ m~s$^{-2}$ and
$432\pm 51$ km~s$^{-1}$ values reported by \citet{crb21},
who also find that the acceleration is over by the time the CME 
reaches the HI1-A field of view.


\section{Morphological Reconstruction}

     The goal in this section is to model the 3-D structure of the
CME, including the internal structure, but we start by modeling the
outline of the overall CME assuming an FR shape.  For this
purpose, we use well-established techniques that have already been
applied extensively in the study of CMEs observed by SOHO, STEREO, and
PSP \citep{bew09,bew17,bew20}.  The construction
of an FR shape begins with the definition of loops that constitute the inner
and outer edges of a 2-D FR in an xy-plane with the x-axis pointing through
the apex of the FR.  In polar coordinates in this plane, these two loops
can be expressed as
\begin{equation}
r(\theta)=r_{max} \exp \left( - \frac{1}{2} \left|
  \frac{\theta}{\sigma} \right| ^{\alpha_s} \right).
\end{equation}
The $\sigma$ parameter determines the widths of the loops, which can
also be described by a full-width-at-half-maximum,
\begin{equation}
FWHM_s=2\left( 1.386 \right)^{1/\alpha_s} \sigma.
\end{equation}
The $\alpha_s$ parameter controls the shape of the top of the loop.

     We have to define the electron density distribution for the 2-D FR,
as the ultimate goal is to create synthetic images of the CME from a
density cube containing the final 3-D FR shape.  In the past, we have
placed mass only on the surface of the FR, leaving the interior empty,
as we were only interested in the outline of the CME shape.
However, we are here interested in the internal structure as well, so
we change our approach.  We start by assuming a uniform density inside
the FR, $n_1(x,y)=n_{FR}$.  The densities of the internal structures described
below will be defined relative to this baseline density.  The absolute
value of $n_{FR}$ is actually unimportant for our purposes, as we are only
comparing the general appearance of the CME structure in the real and
synthetic images, and we are not attempting to make our synthetic images
match observed brightness values quantitatively.

     The density is then made distance dependent, assuming
\begin{equation}
n_2(x,y)=n_1(x,y) \left( \frac{x}{x_{max}} \right) ^{\beta},
\end{equation}
where $x_{max}$ is the distance to the leading edge of the FR from
Sun-center along the x-axis \citep{bew09,bew10}.
If we maintained constant density (e.g., $\beta=0$), the legs of the model
CME would end up looking brighter in synthetic images than the top of the CME,
since the legs are closer to the Sun and therefore scatter more light.
This is not consistent with the appearance of CMEs in images.  We correct
for this by assuming that density increases with distance.  Specifically, we
assume $\beta=3$, although the precise value of $\beta$ is relatively
unimportant, since as already noted we are not attempting to quantitatively
match observed brightness values.

     With the 2-D FR and its density map established, the two loops
of the FR are then used to define a 3-D FR shape and density distribution
by assuming a circular cross section for the FR, bounded by the two loops.
By stretching the FR in the direction perpendicular to the FR creation
plane, i.e.\ the z-axis, an FR can be created with an arbitrary
ellipticity, with density independent of z.  The 3-D FR
is then rotated into the desired orientation in an HEE coordinate system.
Adjusting the various quantities involved in the FR creation process
allows experimentation with different shapes and orientations.

\begin{table}[t]
\small
\begin{center}
Table 1:  CME Morphological Parameters
\begin{tabular}{clccc} \hline \hline
Parameter & Description & FR & Loop 1 & Loop 2 \\
\hline
$\lambda_s$ (deg)& Trajectory longitude & -57   & -57   & -57  \\
$\beta_s$ (deg)  & Trajectory latitude  &  10   &  10   &  10  \\
$\gamma_s$ (deg) & Tilt angle           &   8   & ...   & ...  \\
FWHM$_s$ (deg)   & Angular width        & 48.4  & 30.4  & 19.4 \\
$\Lambda_s$      & Aspect ratio         & 0.096 & 0.027 & 0.027\\ 
$\eta_s$         & Ellipticity          & 1.0   & 1.0   & 1.0  \\
$\alpha_s$       & Leading Edge Shape   & 2.5   & 4.0   & 4.0  \\
$x_{max}$        & Leading Edge Dist.   & 1.0   & 0.95  & 0.95 \\
\multicolumn{5}{l}{\underline{Parameters for Computing $\gamma_s$}} \\
$\gamma_{base}$ (deg)&Base $\gamma_s$   & ...   & 25    & 25   \\
$\gamma_1$ (deg) &Initial top $\gamma_s$& ...   &  20   & -15  \\
$\gamma_2$ (deg) & Final top $\gamma_s$ & ...   & -20   & -75  \\
$\xi$            & Writhe index         & ...   & 3.5   & 3.5  \\
$t_0$ (hr)       & Midpoint time        & ...   & 12    & 12   \\
$t_r$ (hr)       & Relaxation time      & ...   &  4    &  4   \\
\hline
\end{tabular}
\end{center}
\end{table}
     Once a density cube is prepared, synthetic images are generated
using a white-light rendering routine to perform the necessary calculations
of Thomson scattering within the density cube \citep{deb66,afrt06}.
The FR is assumed to expand in a self-similar
expansion, meaning its shape does not change with time, although this
will not be strictly true for the internal structure described below.  The
expansion of the CME is described by the kinematic model in Figure~6.  In
this way, we compute synthetic images of the CME for comparison with the
images from STEREO-A, PSP/WISPR, and SOHO/LASCO.  Parameters of the FR are
adjusted to maximize agreement between the real and synthetic images.
This is done by trial-and-error, and subjective judgment is used to decide
what is the best fit.  The primary parameters of interest are listed in
Table~1.  Figure~7 shows the final FR shape that we decide
best describes the outline of the CME, and Figures 2-5 show synthetic images
computed from this FR for comparison with the real images.
\begin{figure}[t]
\plotfiddle{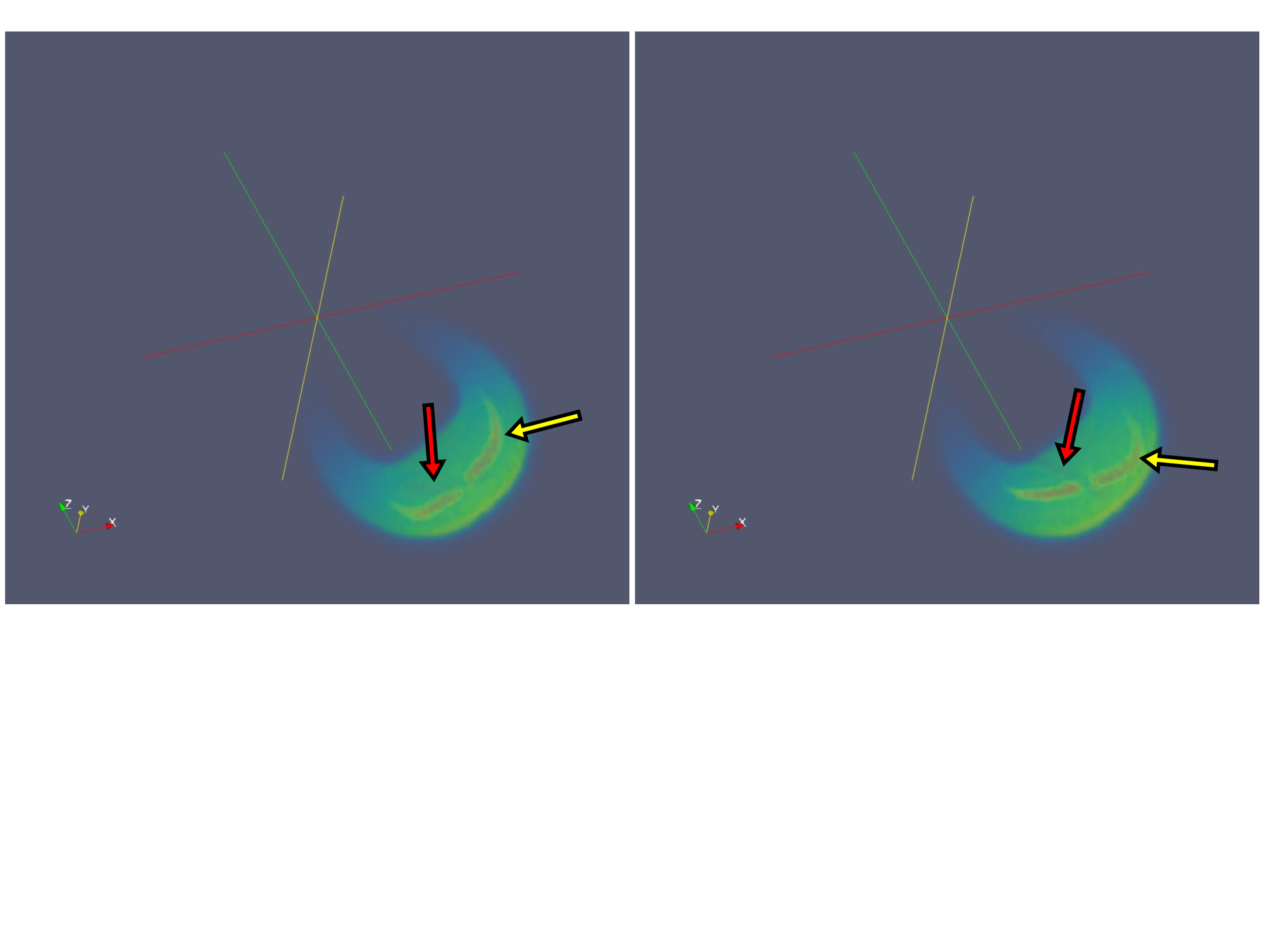}{2.9in}{0}{64}{64}{-230}{-130}
\caption{Reconstructed 3-D FR structure of the CME observed by
  {\em PSP}/WISPR on 2019 April 2, shown in HEE coordinates.  The
  reconstruction includes two loops inside the FR, the high density
  Loop 1 (yellow arrow) and the low density Loop 2 (red
  arrow).  The loops rotate with time, and
  the right and left panels show the structure at the beginning
  and end of the rotation, respectively.}
\end{figure}

     The parameters in Table~1 defining the shape of the
FR use the variable names from \citet{bew17}.  Briefly,
$\lambda_s$ and $\beta_s$ are the central trajectory in HEE
coordinates, with the $\lambda_s=-57^{\circ}$ direction explicitly
indicated in Figure~1.  The FR trajectory latitude is slightly above the
ecliptic, with $\beta_s=10^{\circ}$.
The $\gamma_s$ parameter indicates the tilt angle of the FR,
with $\gamma_s=0^{\circ}$ corresponding to an E-W orientation parallel
to the ecliptic, and $\gamma_s>0^{\circ}$ indicating an upward tilt of
the western leg.  With $\gamma_s=8^{\circ}$, our FR is close to being
oriented E-W.  The FWHM$_s$ parameter is the full-width-at-half-maximum
angular width of the FR, defined by Eqn.\ (2).  The aspect ratio,
$\Lambda_s$, indicates
the radius of the apex of the FR divided by the distance of the apex
from the Sun, and therefore quantifies the thinness of the FR.  
Our FR reconstruction scheme allows for the possibility of an
elliptical FR channel, but we see no evidence for ellipticity, so
$\eta_s=1.0$.  Finally, the $\alpha_s$ parameter from Eqn.\ (1) defines
the shape of the FR leading edge \citep[see][]{bew09,bew17},
with higher values leading to flatter leading edges.
     
     With the overall FR shape determined, we now turn our attention
to the internal structure, namely the bright and dark linear features
noted in Section~2.  We interpret these features as being narrow loop-like
structures within the CME, and we attempt to estimate their shapes from
the WISPR-I images.  We will henceforth refer to the bright and dark
loops as Loop 1 and Loop 2, respectively.  Since the FR creation
procedures described above are already designed to create loop-like
structures, we can apply these same routines to modeling Loops 1 and 2.
Our best-fit loop parameters are therefore listed in Table~1 alongside
the FR parameters.  The FR and two internal loops structures
defined by the Table~1 parameters are ultimately combined into a single
density cube before the white light rendering calculations are made.
The internal loops naturally end up being much thinner (i.e.,
very low $\Lambda_s$) than any CME FRs that these routines have been
used to model in the past.  In order to make Loop 1 bright and Loop 2
dark, within the boundary of Loop 1 we simply increase the ambient
density within the FR by a factor of 3, and within the boundary of Loop 2
we simply set the density to zero.  The meaning and implications
of this density contrast will be addressed in more detail in
Sections 5 and 6.

     The constraints on the loop properties are more limited than for the
FR as a whole, because we only have the WISPR-I images to go by, while
for the FR we have the STEREO-A and SOHO/LASCO images as well.  One
additional constraint is that we naturally require the loops to reside
within the CME FR shape that has already been constructed.  Thus, their
trajectory directions are assumed identical to the FR.  With this
constraint, we find we can only come close to reproducing the observations
if the loops are writhed, i.e.\ with different rotation at their tops
than at their bases.  In the mathematical framework described above, this
is equivalent to saying that the tilt angle parameter, $\gamma_s$,
is dependent on distance from the Sun, i.e.\ $\gamma_s=\gamma_s(x)$.
We assume the following parametrized functional form for $\gamma_s$:
\begin{equation}
\gamma_s(x)=\gamma_{base}+(\gamma_{top}-\gamma_{base})
  \left( \frac{x}{x_{max}} \right) ^{\xi},
\end{equation}
where $\gamma_{base}$ and $\gamma_{top}$ are the $\gamma_s$ values at the
loop base and top, respectively, and $\xi$ is an additional power law
index parameter that affects the degree to which the writhe is focused
at the top of the loop.

     An additional complication related to $\gamma_s$ for the two loops
is that we believe the WISPR-I movies of the CME indicate rotation of
the tops of the loops during their passage through the WISPR-I field of view,
particularly the bright Loop 1.  Some sense of this can be seen by
comparing the two images in Figure 2.  This means that $\gamma_s$ and
$\gamma_{top}$ are actually time-dependent.  In order to model this, we assume
\begin{equation}
\gamma_{top}(t) = \gamma_1 + \left[ \frac{\gamma_2-\gamma_1}{2} \right]
  \left[ \tanh{\left( \frac{t-t_0}{2t_r} \right) } + 1 \right],
\end{equation}
where $\gamma_1$ and $\gamma_2$ are the initial and final values of
$\gamma_{top}$, respectively.  The time $t_0$ is a midpoint time where
$\gamma_{top}$ is halfway between $\gamma_1$ and $\gamma_2$, and the
relaxation time $t_r$ defines how rapidly the shift from $\gamma_1$ to
$\gamma_2$ occurs.  Note that we have used an expression of
time-dependence similar to Eqn.~(6) in modeling a CME from
2008 June 1 \citep{bew10}.  In this case, we do not perceive
any clear time dependence for the FR itself, only for the two
internal loops inside it.

     We use trial-and-error to experiment with the
various loop parameters to find the best match for the data.
The resulting parameters are listed in Table 1.  For the loops, the
$\gamma_s$ parameter is space and time-dependent, and is therefore
replaced with the six parameters listed at the bottom of the
table.  For the time parameters, the $t=0$ time is the same as in
Figure 6, UT 03:06 on 2019 April 2.  Thus, the $t_0=12$~hr time
found for both loops corresponds to UT 15:06 on 2019 April 2,
which is roughly the time of the first WISPR-I image in Figure 2.
A loop parameter not yet mentioned is one that relates
the size of the loops relative to the existing FR.  This is quantified
by $x_{max}$ in Table~1, which is the loop-top distance divided by
the leading edge FR distance, which is $x_{max}=0.95$ for both loops.

     The two panels of Figure 7 show the appearance of the two loops
within the FR at early and late times, i.e.\ with
$\gamma_{top}=\gamma_1$ and $\gamma_{top}=\gamma_2$, respectively,
indicating the degree of rotation that we perceive for the loops.
Figure 2 indicates the degree to which the two loop model is able to
reproduce the observed internal structure.  The agreement is
imperfect, but the synthetic images capture the general characteristics
of the CME's appearance well enough to support the two loop
interpretation of the CME.  Improving the agreement would probably
require the use of more complex and asymmetric shapes than are
possible with the parametrized forms used here.  In the 1 au images
in Figures 4-5, the internal CME structure is difficult
to make out even in the synthetic images, particularly the dark Loop 2,
consistent with our inability to see this structure in the actual images.
This is in part an issue of signal-to-noise, as this is not
a particularly bright CME as viewed from 1~au, but this is also due to
blurring of the very narrow loops, caused by intrinsic resolution and by
lengthy exposure times in HI1 and HI2 compared with WISPR-I data.

\section{The Origins of the Time Dependence}

     The cause of the rotation of Loops 1 and 2 is worthy of
more discussion.  We were initially concerned that this could be an
optical illusion caused by the changing perspective of the CME as it
passes through the WISPR-I field of view.  Ironically, an excellent
example of this effect is apparent in the movie of synthetic HI2-A images
of the CME (see movie version of Figure 4).  The bright Loop 1
is visible in the synthetic images (though not in the real
images) and appears to be rotating as it moves through the HI2-A field
of view.  But this rotation is entirely illusory, because in the model
described above, the loop rotation ceases long before the CME enters the
HI2-A field of view.  The apparent rotation is due to the upper
leg of the loop having a faster apparent motion through the field of view
than the lower leg due to the upper leg's closer proximity to STEREO-A.
This projection effect combined with the ``fish-eye'' distortion of the
very large HI2-A field of view leads to the erroneous impression of
rotation.  We tried many static loop shapes to see if we could induce
this effect in the WISPR-I field of view, but we were unable to find
a shape that would produce it.  The WISPR-I field of view is not
as large as HI2-A, and the image distortions are not as strong.
Thus, we conclude that the rotation is likely real.

     We hypothesize that the cause of the rotation might be forcing
on the CME from behind by a faster ambient solar wind.  This wind is
most apparent in the COR2-A images (see movie version of Figure 4),
where a fast flow is seen behind the CME, which is in fact faster than
the leading edge of the CME at that point.  One apparent
consequence of this is that there is a pileup of material on the
back side of the FR channel, which makes the trailing edge of the
FR channel brighter than the leading edge in the STEREO-A images,
as marked by a yellow arrow in Figure 4.  In HI1-A, the trailing edge
has a distinct V-shaped appearance.  A final effect of external
forcing on the CME may be a degree of pancaking, seen
late in the HI1-A part of the Figure~4 movie and into the HI2-A field
of view.  Such flattening of CMEs relative to the direction of propagation
has been studied previously \citep[e.g.,][]{nps11,ck21}.  This could
in principle be considered in our modeling by making the ellipticity
of the FR channel ($\eta_s$) time dependent, allowing it to increase with
time.  However, this is happening well after the CME is in the WISPR-I field
of view, and is therefore outside the focus of our study.

     The usual conception of a CME is of a fast eruption
of material plowing through a slower ambient solar wind, but the
situation can be reversed for a slow CME like the 2019 April 2 event.
This forcing from behind could be responsible for inducing the
observed rotation in the internal CME structure,
particularly if some of the fast flow seen behind the CME is actually
going up the legs of the FR.  It is also possible that the brightness of
Loop 1 might be due to mass loading into the loop from this fast flow
from behind.

     The origin of the fast wind could be AR12737, which we have
also associated with the origin of the CME (see Section 2).
\citet{lh21} and \citet{dhb21} have studied this
active region extensively with Hinode's EUV Imaging Spectrometer (EIS),
focusing on a region of strong blueshifted Fe XII emission on the east
side of the active region.  They identify this outflow region as
the likely source of Type III radio bursts observed by PSP.  Such
outflow regions beside active regions have been interpreted as
being possible sources of solar wind.  As such, it is possible
that the fast wind flow observed behind the April 2 CME in COR2-A
is connected to this outflow region.

\section{Interpreting the Internal Structure}

     We have interpreted the internal structure of the 2019 April 2 CME
seen by WISPR-I as being due to two time-dependent writhed loops, one
bright (Loop 1) and one dark (Loop 2).  The next step is to address the
question of what the existence of these loops implies about the nature of
CME structure.  On this issue there is still significant ambiguity.

     We have placed the loops within a CME outline defined by an FR
shape.  The magnetic FR paradigm has become the predominant paradigm
for CME field structure, with support from both in situ and imaging
data \citep{rpl90,cjf95,jc97,vb98,av13,av14,bew17}.
Within this paradigm, our two loops might be
interpreted as indicative of field lines within the FR.  This
is potentially problematic, because we are not necessarily perceiving
the field lines to be twisted around a current axis in a helical fashion,
as one would expect within the conventional FR picture \citep[for further
discussion see][]{sp20}.


     However, it seems simplistic to interpret Loops 1 and 2 as simple magnetic
field lines, as their lateral widths are clearly resolved and too large to
represent single field lines.  They must instead be collections of field lines.
Another interpretation of the loops is that they are
in fact separate and independent FRs within the CME outline (or helical field
bundles, at least).  Support for this
interpretation is provided by the many instances of in situ observations of
CMEs implying the existence of multiple FRs \citep{vao99,qh03,cjf11,qh21},
including one of the first CMEs observed in situ by PSP \citep{tnc20}.
Surveys of ``magnetic clouds'' (MCs) identified in in situ data, such as
\citet{rpl11,rpl15}, include numerous instances of MCs close together
in time, which could be FRs within the same CME.  Of the 28 CMEs studied
by \citet{bew17}, there were three cases that yielded two MCs in the
\citet{rpl11,rpl15} lists instead of just one, suggesting that
roughly 10\% of CMEs are perceived as having multiple FRs when they happen
to encounter a spacecraft.

     It is possible that most CMEs actually contain multiple FRs, potentially
arising from interactions with ambient flux systems during CME eruption or
propagation.  Although
most spacecraft encounters with CMEs do not perceive this, we would
not actually expect that they would, since a single path through a CME will
only sample a small part of it.  For example, a random track through the
reconstruction shown in Figure 7 would be unlikely to hit both Loops 1 and 2
squarely, making it unlikely that a spacecraft encounter with the 2019 April 2
would have been able to discern the presence of both structures.  This could
help explain the inconsistent FR orientations inferred for certain CMEs that
hit multiple spacecraft \citep{cjf11,cm12}.
The different spacecraft might actually be sampling different FRs within the
same CME.

     The idea that CMEs actually consist of multiple FRs could also resolve
a fundamental size discrepancy that seems to exist between FR structures
inferred from images and those inferred from in situ data.
In a survey of in situ selected events, \citet{bew17} find that
magnetic cloud encounter times observed near Earth are on average three
times shorter than expected based on FR reconstructions from images.
If CMEs usually consist of two or more FRs, the CME shape
inferred from the images would represent the outline of the combined
collection of FRs, while the in situ data would typically only be encountering
one of the FRs, thereby explaining the inconsistency.

     It would be worthwhile to try to find other CMEs that have internal
structure that might also be interpreted as being indicative of multiple
FRs, both with WISPR and with white light imagers operating at 1 au.
However, the dearth of prior CME studies suggesting such an interpretation
suggests that the 2019 April 2 CME is unusual in this respect.  We suspect
that what makes it different is the unusual mass loading that seems to be
occurring in Loop 1, making it bright in the WISPR-I images, relative to
both Loop 2 and the rest of the CME interior.  We have hypothesized that
perhaps the mass loading is occurring because the footpoints of Loop 1 are
at least in part connected to the outflow region to the southeast of AR12737,
while Loop 2 and whatever other magnetic elements might exist within the
CME are rooted elsewhere.  Normally, CME interiors have rather low density, and
if there is no density contrast between the internal magnetic structures
within a CME, there will be no way to see them.
The 2019 April 2 CME may be a rare case with sufficient density contrast
to perceive the structures, and we still would not have been able to see
this without PSP/WISPR being able to view the CME close to the Sun.

\section{Summary}

     We have analyzed white light images of the 2019 April 2 CME
from PSP/WISPR, STEREO-A, and SOHO/LASCO, with a focus on the
transient's internal structure.  Our findings are summarized as
follows:
\begin{description}
\item[1.] The CME appears to be caused by the destabilization of the
  streamer belt by a newly emerged and expanding underlying active
  region, AR12737.
\item[2.] The analysis includes a comprehensive kinematic model for the
  CME, relying mostly on STEREO-A, which is able to track the CME
  continuously from the Sun to 1 au.  Using all available imaging
  data we reconstruct the CME's 3-D morphology assuming an FR shape.
\item[3.] In WISPR-I images we see internal structure within the CME
  that is not apparent in images from the 1 au spacecraft, emphasizing
  the usefulness of being able to observe CMEs close to the Sun.
  We interpret the internal CME structure as consisting of two
  loops within the FR, one bright and one dark.  The loops are
  writhed, and the tops of the loops appear to exhibit rotation within
  the WISPR-I field of view, meaning that the writhe is increasing with
  time.  We model the time-dependent 3-D structure of the two loops
  using methodology similar to that used to model the overall CME
  FR shape.
\item[4.] In the STEREO-A images, we see clear evidence for a fast
  ambient solar wind flow overtaking the CME from behind, resulting
  in a visible pileup of material on the backside of the FR channel.
  We speculate that some of this flow may be going up the legs of the
  CME, and that this could be responsible for the mass loading that
  makes Loop 1 bright, and also for the increase in writhe of the
  two loops.  This fast flow might be connected to an outflow region
  observed by Hinode southeast of AR12737.
\item[5.] The two-loop reconstruction of the CME's internal structure
  remains subject to interpretation, but our favored interpretation
  is that the two loops are separate FRs, consistent with
  the numerous published examples of multiple MCs being
  detected in situ within the same CME.  It is possible that
  most CMEs possess multiple FRs within them,
  but the separate structures are rarely perceived in images due to low
  internal densities or low relative contrast, and they are only occasionally
  perceived in situ when a spacecraft happens to encounter
  more than one of the internal FRs in its track through the CME.
\end{description}

\acknowledgments

     Financial support was provided by the Office of Naval Research.
{\em Parker Solar Probe} was designed, built, and is now operated by the
Johns Hopkins Applied Physics Laboratory as part of NASA's Living with a
Star (LWS) program (contract NNN06AA01C).  We particularly acknowledge
the WISPR instrument team, funded by NASA through grant NNG11EK11I.
The STEREO/SECCHI data are produced by a
consortium of NRL (US), LMSAL (US), NASA/GSFC (US), RAL (UK), UBHAM
(UK), MPS (Germany), CSL (Belgium), IOTA (France), and IAS (France).
In addition to funding by NASA, NRL also received support from the
USAF Space Test Program and ONR.  C.R.B. acknowledged support from
NASA's STEREO/SECCHI project (NNG17PP27I).  A.V. acknowledges support
from NASA grants 80NSSC20K1282 and 80NSSC19K1261.

\end{document}